# Fast 3D $^{31}$P $B_1^+$ mapping with a weighted stack of spiral trajectory at 7 Tesla

Mark Widmaier[1,2,3], Antonia Kaiser[1,2], Salome Baup[1,2], Daniel Wenz[1,2], Katarzyna Pierzchala[1,2], Ying Xiao[1,2,3], Zhiwei Huang[1,2], Yun Jiang[5], Lijing Xin[1,2,4*]

1. CIBM Center for Biomedical Imaging, Switzerland
2. Animal Imaging and Technology, Ecole Polytechnique Federale de Lausanne (EPFL), Lausanne, Switzerland
3. Laboratory of functional and metabolic imaging, Ecole Polytechnique Federale de Lausanne (EPFL), Lausanne, Switzerland
4. Institute of Physics, Ecole Polytechnique Federale de Lausanne (EPFL), Lausanne, Switzerland
5. Department of Radiology, University of Michigan, Ann Arbor, Michigan, USA

*Corresponding author:
Lijing Xin, Ph.D., EPFL AVP CP CIBM-AIT, Station 6, CH-1015 Lausanne, Switzerland
lijing.xin@epfl.ch
Tel.: + 41 21 693 0597

**Word count: 4303**

**Abstract:**

**Purpose:** Phosphorus Magnetic Resonance Spectroscopy ($^{31}$P MRS) enables non-invasive assessment of energy metabolism, yet its application is hindered by sensitivity limitations. To overcome this, often high magnetic fields are used, leading to challenges such as spatial $B_1^+$ inhomogeneity and therefore the need for accurate flip angle determination in accelerated acquisitions with short repetition times ($T_R$). In response to these challenges, we propose a novel short $T_R$ and look-up table-based Double-Angle Method for fast 3D $^{31}$P $B_1^+$ mapping (fDAM).

**Methods:** Our method incorporates 3D weighted stack of spiral gradient echo acquisitions and a frequency-selective pulse to enable efficient $B_1^+$ mapping based on the phosphocreatine signal at 7T. Protocols were optimised using simulations and validated through phantom experiments. The method was validated in phantom experiments and skeletal muscle applications using a birdcage $^1$H/$^{31}$P volume coil.

**Results:** The results of fDAM were compared to the classical DAM (cDAM). A good correlation (r=0.94) was obtained between the two $B_1^+$ maps. A 3D $^{31}$P $B_1^+$ mapping in the human calf muscle was achieved in about 10 min using a birdcage volume coil, with a 20% extended coverage relative to that of the cDAM (24 min). fDAM also enabled the first full brain coverage $^{31}$P 3D $B_1^+$ mapping in approx. 10 min using a 1 Tx/ 32 Rx coil.

**Conclusion:** fDAM is an efficient method for $^{31}$P 3D $B_1^+$ mapping, showing promise for future applications in rapid $^{31}$P MRSI.

**Introduction:**

Phosphorus ($^{31}$P) Magnetic Resonance Spectroscopic Imaging (MRSI) enables in vivo probing of energy metabolism[1,2]. Although it was demonstrated that $^{31}$P MRSI can be a useful method to study e.g. cancer[3], neuropsychiatric[4,5] and neurodegenerative diseases[6–8], its clinical application is constrained by long acquisition times resulting from inherently low $^{31}$P NMR sensitivity. To partially address this issue, ultra-high magnetic fields (UHF), such as 7T, are used[1,9–12]. However, higher $B_0$ results in a less homogenous transmit field ($B_1^+$)[13,14], and consequentially in a higher uncertainty of the estimated flip angle (FA) distribution. The FA distribution is critical to assess the concentration changes of different metabolites using $^{31}$P MRSI in a reliable manner[12,15–21]. For this purpose, in vivo $B_1^+$-mapping, which can be quite time-consuming, is performed. Therefore, it is of highest relevance to reduce the acquisition time needed for this step.

Various $B_1^+$ mapping methods for $^1$H MRI have previously been proposed, including the Double-Angle Method (DAM)[22,23], SA2RAGE[24], saturated DAM[25], actual flip angle method (AFI)[26], and phase-dependent methods[27,28] such as Bloch Siegert (BS) shift method[29–31]. However, their direct translation to $^{31}$P or other X-nuclei $B_1^+$ mapping is challenging due to differences in sensitivity, relaxation properties, and deviating resonance peaks at various chemical shifts. Initial adaptations of some of these methods have been successfully applied in $^{31}$P $B_1^+$ mapping, such as DAM[32], AFI[15], and BS[33].

Nevertheless, these methods exhibit several drawbacks. DAM, from now on referred to as classical DAM (cDAM), is often considered as the standard reference method, but is hampered by prolonged acquisition times due to the necessity of long $T_R$ ($5T_1$) required for a fully relaxed condition. Especially $^{31}$P metabolites exhibit extended $T_1$ relaxation times, leading to prolonged acquisition times, limited spatial resolution, or potential errors when using reduced TRs. Short $T_R$ DAM methods have been proposed for $^1$H, relying on a steady state acquisition. However, the method of Ishimori et al. is limited to a range close to the nominal flip angle, leading to bias in the $B_1$ estimation, and has not been demonstrated in vivo[34]. The other method from Bouhrara et al. is challenging to apply in X-Nuclei imaging suffering from low SNR with proposed flip angles[25]. Other methods, as the recently proposed 2D dual-TR[35], the AFI[15] method and a 3D BS-based method[33] are also operating in steady state to reduce the acquisition time. Even though the BS method has proven to be a promising method as it is independent of $T_1$ relaxation times, it needs a dedicated sequence, with an additional Fermi pulse. This pulse introduces additional RF power deposition and prolonged $T_E$, which might lead to sensitivity loss due to weighting in the effective transversal relaxation time. In the dual-TR and BS method, the localisation is based on a chemical shift imaging (CSI) readout, which limits the spatial and temporal resolution. In a different work, a $^{31}$P AFI MRSI approach[15] was implemented using the phosphocreatine (PCr) signal by applying a frequency-selective pulse. However, the approximation of a much shorter $T_{R,1}$ and $T_{R,2}$ compared to $T_1$ comes at costs of SNR with a flip angle (FA) of 60°[15,26] or if $T_{R,2}$ is close to $T_1$ with a bias in the $B_1^+$ estimation. In all presented works, validation of the method was only performed by surface coils to overcome the low signal-to-noise ratio (SNR), which limits the application to a small surface area.

To address these problems, we propose a short $T_R$ and look-up table based fast DAM (fDAM) for time-efficient $^{31}$P $B_1^+$ mapping. 3D weighted stack of spiral gradient echo (GRE) acquisitions and a frequency selective pulse were incorporated to enable an efficient $^{31}$P $B_1^+$-mapping approach based on the PCr signal at 7T using a $^{31}$P/$^1$H birdcage volume coil and a $^{31}$P/$^1$H head coil array.

## Methods:

### *Parameter optimisation*

The cDAM involves the acquisition of two images, one with flip angle (FA) $\alpha_1$ and the other with FA $\alpha_2$[22,23]. The achieved FA $\alpha_e$ can be estimated by

$$\alpha_e = \mathrm{acos}\left(S_{\alpha 2}/2S_{\alpha 1}\right). \qquad (1)$$

This formula can be derived for the asymptotic value of $T_R \to \infty$. The signal intensity $S_\alpha$ of one image obtained using a GRE sequence is hereby described as

$$S_\alpha = K\rho \frac{(1 - \exp(-\sigma))\sin(\alpha)}{1 - \cos(\alpha)\exp(-\sigma)} \exp\left(\frac{-T_E}{T_2^*}\right), \qquad (2)$$

where $K$ is a scaling factor, $\rho$ is the spin density, $T_2^*$ the apparent relaxation time, $\sigma = T_R/T_1$, and $T_E$ is the echo time. The normalised signal intensity ratio $r$ between the two GRE images can be described as

$$r = \frac{\alpha_1 S_{\alpha 2}}{\alpha_2 S_{\alpha 1}} = \frac{\alpha_1 \sin(C_{B1}\alpha_2)}{\alpha_2 \sin(C_{B1}\alpha_1)} \frac{(1 - \cos(C_{B1}\alpha_1)\exp(-\sigma))}{(1 - \cos(C_{B1}\alpha_2)\exp(-\sigma))}, \qquad (3)$$

where $C_{B1}$ is the unitless $B_1^+$ scaling factor and can be interpreted as the scaling between the actual and nominal flip angle. $C_{B1}$ is used as metric in this work to interpretate the results. $C_{B1}$ can be transformed to $B_1^+$ values, using the hard pulse equivalent of the same nominal α_1 in radians by

$$B_1^+ = \frac{C_{B1} \cdot \alpha_{1,\text{rad}} \cdot 10^6}{0.431 \cdot T_p \cdot \gamma} \ [\mu T], \tag{4}$$

where $T_p$ is the pulse length, $\gamma$ the gyromagnetic ratio and the factor 0.431 is the hard pulse transformation factor for the Gaussian pulse used in this work. As $T_R$, $\alpha_1$ and $\alpha_2$ are known for a given $T_1$ value, a look-up table can be generated connecting different signal ratios $r$ to the $B_1^+$ scaling factor $C_{B1}$. The factor $\alpha_1/\alpha_2$ in Eq. 3 normalises the look-up table to a maximum value of 1. Figure 1b displays different signal ratios for different $\sigma$ at different actual FAs ($\alpha_1 C_{B1}$). The estimated FA is determined by searching the closest value of the measured signal ratio $r$ in the look-up table. Throughout this work, we set $\alpha_2 = 2\alpha_1$, as in the cDAM approach.

For an optimal $\alpha_1$, three factors are considered in this work: 1) temporal SNR gain $G_{\alpha 1}$ of the GRE for $\alpha_1$; 2) $G_{\alpha 2}$ of the GRE for $\alpha_2$ and 3) the sensitivity of the look-up table. $G_\alpha$ is the temporal SNR gain relative to that of a GRE at fully relaxed condition with $\sigma = 5$ and $\alpha_1 = 90°$:

$$G_\alpha = \frac{SNR_\alpha(\sigma, \alpha)}{SNR_{90°}(5, 90°)} = \frac{S_\alpha(\sigma, \alpha)}{S_{90°}(5, 90)} \cdot \sqrt{\frac{5}{\sigma}}. \tag{5}$$

The sensitivity of the look-up table is defined as

$$s = \alpha_1 \left(\frac{dr}{d\alpha_1}\right). \tag{6}$$

The multiplication of $\alpha_1$ hereby scales the derivative to make $s$ independent of $\alpha_1$. As a detected 1° change, choosing $\alpha_1 = 10°$ is a 10% change in $B_1^+$, whereas it is only a 2% change for a $\alpha_1 = 50°$. To achieve the same sensitivity as $\alpha_1 = 50°$ the derivative when choosing $\alpha_1 = 10°$ must be 5 times larger. The values of the objective function $O$ for the optimization are found by a multiplication of all 3 attributes,

$$O = G_{\alpha 1} \cdot G_{\alpha 2} \cdot s \tag{7}$$

and is shown in Figure 1c. The optimal $\alpha_1$ for a given $\sigma$ is found by the maximum value of $O$ and is indicated in Figure 1c. The overall optimal $\alpha_1$ was found at $\alpha_{1,opt} = 59°$ and $\sigma_{opt} = 1.6$.

### *31P MRI sequence and reconstruction*

Phantom and in vivo calf muscle experiments were conducted on a 7T/68 cm MR scanner (Siemens Healthineers, Erlangen, Germany) with a 28 cm diameter [31]P/[1]H circularly polarized (CP) birdcage coil (Stark Contrast, Erlangen, Germany). Additionally, an in vivo brain experiment was performed on a 7T Terra X system (Siemens Healthineers, Erlangen, Germany) with a double tuned Tx/Rx $^{31}$P/$^{1}$H birdcage coil (26 cm diameter) and a 32 channel $^{31}$P-Rx only phased array (Rapid Biomedical, Rimpar, Germany). The 3D spiral $^{31}$P GRE sequence diagram (version VB17 and XA60 available for sequence transfer via c2p agreement) is depicted in the Figure 2. An 8 ms frequency selective Gaussian pulse with a bandwidth of 340Hz was used to excite phosphocreatine (PCr). For the VB17A version a 21.08 ms variable-density spiral trajectory[36] (Figure 2c) after a $T_E = 200$ µs was deployed. Due to gradient stimulation limits the spiral trajectory had to be redesigned for the XA60 version. The slew rate limits were reduced from 150 T/(m s)$^{-1}$ to 100 T/(m s)$^{-1}$. The duration of the non-uniform spiral for the XA60 version is 18.72 ms and a $T_E = 410$ µs (Figure 2d). The encoding along $k_z$ was achieved by a stack of spiral with a Hamming weighted averaging scheme (Figure 2b). One 3D k-space sampling was achieved with a total of 13 $k_z$ encoded spiral readouts. The total acquisition time was, therefore, $TA = 2 \cdot n_{rep} \cdot 13 \cdot T_R$, with $n_{rep}$ the number of repetitions of a 3D volume acquisition for SNR enhancement and the factor 2 as two GRE images must be sampled. The matrix size of the reconstructed image space was $16 \times 16 \times 11$ in phantom and calf muscle and $32 \times 32 \times 11$ in human brain for a $230 \times 230 \times 220$ mm$^3$ field of view (FOV). The vector size of one spiral without the rewinding is $L = 2061$ (VB17A) and $L = 1782$ (XA60), respectively. The uniform spiral trajectories of both versions, with characteristic dense sampling in the k-space centre, provided sufficient k-space coverage for a single shot in-plane sampling, with a mean oversampling ratio of 8 for the VB17A and 1.7 for the XA60 version, respectively. All data was processed using a custom script in MATLAB (The MathWorks, Inc., Natick, Massachusetts, USA). A 1D-FFT was applied along slice dimension before regridding. The density compensation function (DCF) was calculated based on the Voronoi diagram[37–39]. Before regridding, the DCF was multiplied by a low-pass filter (LP)

$$\mathrm{LP}_1(n) = \cos^2\left(\frac{\pi * n}{L}\right) \cdot e^{-\frac{n}{L}}, \quad (8)$$

a combination of a half periodic Hann filter and an exponential filter to reduce the influence of high-frequency components on the image acquired in the phantom and calf muscle. To cope with the lower SNR in the brain, a stronger smoothing filter,

$$\mathrm{LP}_2(n) = \cos^2\left(\frac{\pi * n}{900}\right) \quad (9)$$

was used and the remaining $L - 900$ points were zero filled. The regridding was performed using a Kaiser-Bessel kernel[40,41] to transform the data in to the Cartesian k-space. Lastly, a 2D-FFT transformed the GRE data into the image domain. The data acquired with the 1Tx/32Rx coil was processed for each channel separately and combined using adaptive combine[42] in the image domain. The GRE images are then used for $C_{B1}$ estimations as described above.

*Phantom validation*

The previously described method was validated in a spherical phantom with a diameter of 17 cm, filled with 50 mM Pi solution (Carl Roth GmbH & Co. KG, Karlsruhe, Germany). The longitudinal relaxation time was measured by an inversion recovery method ($T_1 = 7.2$ s). $C_{B1}$ maps were acquired for $\sigma = 2, 1.6, 1, 0.5$ with $\alpha_1 = 63°, 59°, 53°, 40°$ and $T_R = 14.4, 11.52, 7.2, 3.6$ s respectively, to examine the FA optimisation dependent on $\sigma$. The acquisition time per GRE was fixed to 13:26 min including dummy scans (4, 5, 8, 16 RF pulse repetitions respectively). The repetitions of one 3D acquisition for each $\sigma$ were adjusted accordingly with $n_{\text{rep}} = 4, 5, 8, 16$. The reference voltage $V_{ref}$ was set to 250 V. $B_1^+$ maps acquired by the cDAM ($\sigma = 5$, $\alpha_1 = 65°$) were used as reference images ($n_{\text{rep}} = 2$, 16:30 min acquisition time (TA)). To evaluate the difference between the $C_{B1}$ maps obtained by fDAM and cDAM, the difference maps were generated by calculating the percentage difference per voxel in slice 6. For $\sigma_{opt} = 1.6$ the estimation performance was further evaluated by the Band-Altman plot and Pearson correlations. To calculate the SNR in a voxel, the phantom was masked, and the signal intensity of the voxel was divided by the noise. The noise is defined as the standard deviation (STD) over all voxels inside the mask of the difference between the first and second half of the acquired averages[43].

*$T_1$ effect*

The accuracy of the look-up table approach depends on the knowledge of $\sigma$ and thus $T_1$. To investigate the effect of wrong $T_1$ assumptions, simulations were conducted. A look-up table for $\sigma_{opt} = 1.6$ and $\alpha_{1,opt} = 59°$ was calculated, with $C_{B1}$ was ranging from 0.02 to 2.6 (0.02/step). Simulated signals with $C_{B1}$ ranging from 0.1 to 2.0 (0.02/step) and a deviation of $\sigma$ ranging from -50% to 50% (0.1%/step) were processed in the estimation process. The estimation error was calculated as the percentage difference of the estimated $C_{B1}$ from the actual $C_{B1}$. Further, the simulation of $T_1$ effect was validated in the phantom results for $\sigma_{opt} = 1.6$ by altering the assumed $T_1$ in the look-up table to 50%, 75%, 125%, 150% of the actual $T_1$. The error maps were calculated from the relative difference of these $C_{B1}$ maps from those obtained with the actual measured $T_1 = 7.2$ s. The $T_1$ effect is also $C_{B1}$ dependent according to the simulation results. Therefore, two different $C_{B1}$ values were used in the phantom experiments by conducting experiments at two different $V_{ref} = 250$ V and $V_{ref} = 175$ V.

*In vivo validation in the calf muscle*

In vivo data were acquired in the calf muscle of two participants (2 female, 28 and 29 years old), with written informed consent provided. fDAM $C_{B1}$ maps were acquired, and cDAM $C_{B1}$ maps were acquired as reference. For the fDAM method the sequence parameters are set with $\sigma_{opt} = 1.6$, ($T_R = 5.7$ s, assuming $T_1$ of PCr is 3.55 s[10,15]), $\alpha_1 = 59°$, $\alpha_2 = 118°$ and $n_{\text{rep}} = 6$ (TA 15:45 min; including 5 dummy scans for each GRE). The cDAM $C_{B1}$ maps were acquired with $\alpha_1 = 65°$, $n_{\text{rep}} = 3$, $T_R = 18$ s, with a TA of 23:25 min. For both sequences, the transmit voltage was set to 300 V. To investigate a potential TA reduction from the fDAM with $n_{rep} = 6$, subsets of $n_{rep} = 4$ and $n_{rep} = 2$ were taken to generate the $C_{B1}$ maps "Rep. 4" and "Rep. 2", respectively. The scan time of these reduced data sets resemble a TA of 10:50 min and 5:55 min respectively. Data was compared with relative difference maps, where fDAM $C_{B1}$ maps with $n_{rep} = 6$ were taken as references. Due to the low sensitivity of the birdcage coil in the peripheral regions, only voxels with an SNR > 3 were considered (Monte Carlo simulation in the supplementary files: Figure S1). The acquired maps were then interpolated to $32 \times 32 \times 11$ and masked with accordingly down sampled ¹H GRE images ($128 \times 128 \times 11$, $T_E = 3.37$ ms, $T_R = 15$ ms).

*In vivo validation in the human brain*

In addition to the calf muscle data, $C_{B1}$ maps in the brain were acquired by fDAM from one participant (male, 43 years old, written informed consent provided). The sequence parameters of the fDAM were adapted to the brain with T$_R$=5.4 s, (assuming the $T_1$ of PCr is 3.4 s[9,44]), $\alpha_1 = 59°$, $\alpha_2 = 118°$ ($n_{\text{rep}} = 6$; TA 15:00 min; including 5 dummy scans for each GRE). Reference cDAM $C_{B1}$ maps were acquired with $\alpha_1 = 65°$, $n_{\text{rep}} = 3$, $T_R = 18$ s and a TA of 23:25 min. For both sequences the transmit voltage was set to 400 V. A potential TA reduction from the fDAM with $n_{\text{rep}} = 6$ was investigated as described for the calf muscle. The scan time of the subsamples resemble a TA of 10:15 min for $n_{\text{rep}} = 4$ and 5:35 min for $n_{\text{rep}} = 2$. The acquired maps were masked with manually created brain masks from ¹H GRE images ($128 \times 128 \times 11$, $T_E = 3.37$ ms, $T_R = 15$ ms).

## Results

*Simulation results*

The effect of $\sigma < 5$ using the cDAM (Equation 1) is shown in Figure 1a. As $\sigma$ decreases, FAs are more likely to be overestimated, identifying the need for a correction method. Figure 1b illustrates our look-up table approach, successfully connecting the signal ratio r to an actual FA $\alpha = C_{B1} \cdot \alpha_1$ for a given $\sigma$. To investigate the optimal parameters, values of the optimisation function were plotted as a function of $\alpha_1$ and $\sigma$ (Figure 1c). With the respective $\sigma$, an optimal $\alpha_1$ was found (the grey line in Figure 1c). The highest value of the optimisation function was found at $\sigma_{opt} = 1.6$ (white squared dot in Figure 1c) with $\alpha_{1,opt} = 59°$.

*Phantom Validation*

The results of the phantom validation are displayed in Figure 3. Figure 3a demonstrates the $C_{B1}$ maps of cDAM and fDAM with different $\sigma$ (2.0, 1.6, 1.0, 0.5 for fDAM) and the corresponding $C_{B1}$ relative difference (Diff.) [%] maps between cDAM and fDAM. The $C_{B1}$ maps are in good agreement as indicated by the difference plots which are not exceeding 10% (Figure 3b). Note that the difference is lower in the centre of the slice where the SNR and $C_{B1}$ are increased. With the optimal $\alpha_{1,opt} = 59°$ and $\sigma_{opt} = 1.6$, fDAM was compared over all slices with cDAM using voxel-wise Pearson correlations (Figure 3c) and the mean differences were compared using a Bland-Altman plot (Figure 3d). The estimated $C_{B1}$ obtained by both methods were in excellent agreement with a Pearson correlation coefficient of 0.95. The Bland-Altman plot showed that fDAM demonstrates a small mean overestimation of 0.01 and 95% limits of agreement (1.96 STD) of ± 0.05.

*$T_1$ effect*

Figure 4a illustrates the simulation results of the influence of $T_1$ values on the $C_{B1}$ estimation error [%] using $\alpha_{1,opt} = 59°$ and $\sigma_{opt} = 1.6$. For $C_{B1} > 1$, the estimation error was negligible (<10%). For $C_{B1} < 1$, $C_{B1}$ is overestimated when $T_1$ is underestimated and vice versa. Overall, the $C_{B1}$ bias was within ±25% with a ±50% deviation in $T_1$ and within ±15% with a ±25% deviation in $T_1$.

To validate the impact of $T_1$ on $C_{B1}$ estimation, additional phantom experiments using the optimal $\alpha_{1,opt} = 59°$ and $\sigma_{opt} = 1.6$ were performed with two sets of reference voltages (175V and 250V). The $C_{B1}$ values were estimated using look-up tables simulated with biased $T_1$ values of ±25% and ± 50%. The $C_{B1}$ estimation errors relative to the cDAM with the correct $T_1$ are shown in Figure 4b. A voxel was picked to compare the phantom results with the simulations (Figure 4c). For the reference voltage of 250V, $C_{B1}$ is estimated to be 1.15. Both results indicate that the bias does not exceed 10% in a ±50% deviation range of $T_1$. For the reference voltage of 175V, $C_{B1}$ is estimated to be 0.81 and the bias does not exceed 17% in a ±50% range and 10% in a ±25% deviation range of $T_1$. The error maps are in good agreement with the simulation results. Note that with smaller $C_{B1}$, the estimation errors are more sensitive to $T_1$ bias.

*In vivo validation in the calf muscle*

¹H GRE images, ³¹P GRE images with $\alpha_1$ and $\alpha_2$ in a human calf muscle (left and right leg), cDAM and fDAM $C_{B1}$ maps, as well as their difference maps, are shown in Figure 5. Slices 3 to 10 are shown due to missing coil sensitivity coverage in slice 1, 2 and 11. cDAM and fDAM showed comparable results, especially in the centre slices, where the coil sensitivity is high. fDAM demonstrated an overall higher SNR coverage (SNR >3) compared to cDAM. In most regions, the deviation between the two methods did not exceed 15%. The regions with large difference had lower signals, as seen in the ³¹P GRE images.

The influence of SNR on $C_{B1}$ estimation was evaluated in Figure 6. When decreasing the number of repetitions (Rep.), the overall image coverage was reduced. fDAM with $n_{\text{rep}} = 4$ showed at least 20% more coverage as cDAM in just 40% of the TA of cDAM. Regions with low signal sensitivity demonstrated a larger difference. Using just $n_{\text{rep}} = 2$ (6 min), a good estimation of $C_{B1}$ could be achieved in most regions in the centre slices with deviation below 15% from that measured with $n_{\text{rep}} = 6$.

### *In vivo results in the human brain*

The ¹H GRE images, ³¹P GRE images with $\alpha_1$ and $\alpha_2$, the $C_{B1}$ maps of fDAM, cDAM and fDAM with $n_{\text{rep}} = 4$ and $n_{\text{rep}} = 2$ are shown in Figure 7. Their respective difference maps, taking fDAM with $n_{\text{rep}} = 6$ as reference, are in the line below the $C_{B1}$ maps. Slices 9 to 11 are not displayed as no sensitive tissue is covered in these slices. cDAM and fDAM show comparable $C_{B1}$ maps demonstrated in the difference maps with < 20% in most regions. $n_{\text{rep}} = 4$ is in good agreement not exceeding a difference of 20%, indicating a possible scan time reduction to 11:15 min. With further reduction to Rep. 2 (5:35 min), differences in the $C_{B1}$ maps are increasing in number and severity, however, still doesn't exceed 20% in most regions.

## Discussion:

In this work, we demonstrated the effectiveness of fDAM, a short $T_R$ double-angle $B_1^+$ mapping approach with a look-up table to compensate for $T_1$ saturation effects. An optimised $T_R = 1.6 T_1$ and $\alpha_1 = 59°$ was found by applying an optimisation function. The fDAM was combined with an efficient weighted stack of spiral acquisition and a frequency selective pulse for fast 3D acquisition. The method performance was evaluated in a phantom and the results showed a good correlation (r=0.94) with those of cDAM. For the first time, using a birdcage coil, a 3D $C_{B1}$-map in the human calf muscle was achieved in 10:30 min, with a 20% extended coverage relative to that of the cDAM (23:30 min). Additionally, fDAM enabled the first full brain coverage ³¹P 3D $C_{B1}$-map at 7T in just 10:15 min using a 1 channel Tx/ 32 channel Rx coil. The $C_{B1}$ maps showed a typical birdcage coil behavior at high fields with higher $B_1$ values in the center than the peripheral. For GRE1, the mean $B_1^+$ was 16 µT, 15 µT and 14 µT for participant 1 (calf muscle), participant 2 (calf muscle), and participant 3 (brain), respectively. The related histograms are shown in the supplementary files (Figure S2). The two coils showed differences in transmit efficiency, as the brain experiment was driven with a 33% higher transmit voltage than the muscle experiment. This could attribute to the degraded transmit efficiency due to the accommodation of the 32-channel receive arrays. The reported values and performance[45] stayed in agreement with previous reports of a whole body coil (10.4 µT)[46].

CSI is commonly used as a technique for $^{31}$P $B_1^+$ mapping, offering extra chemical shift information, yet impacting the time efficiency of $B_1^+$ mapping. Our proposed method uses a frequency selective pulse combined with the hamming weighted stack of spiral readout, which results in an efficient acquisition. The resulting temporal SNR benefit allows for a higher (relative to DAM[32], dual-TR[35], BS[33]) or similar (to AFI[15]) spatial resolution in a shorter TA relative to existing methods. Note that previous methods have been demonstrated on surface or quadrature coils known for their high local sensitivity, whereas fDAM is demonstrated in the calf muscle on a less sensitive $^{31}$P/$^{1}$H single-channel birdcage coil. The short acquisition time of fDAM enables its integration in the scanning protocol to obtain a subject-specific $C_{B1}$ map. In this work, we demonstrate the method for the $^{31}$P nucleus, while the look-up table approach and optimal parameters can also be applied to other nuclei and GRE types of sequences with fast readout trajectory.

As brain PCr concentration is by a factor 8 lower than in the calf muscle[9,16], the sensitivity is much lower, making $B_1^+$ map measurement even more challenging. In this work, we took use of two countermeasures: 1) a 1 Tx/ 32 Rx coil to increase the sensitivity and 2) an LP with a lower cut off-frequency ($LP_2$). Note that such LP comes with additional smoothing, which reduces the actual resolution. However, $B_1^+$ field changes smoothly across space, rarely giving the necessity for high resolution $C_{B1}$ measurements.

GRE sequences are known to be sensitive to $B_0$ inhomogeneities[47], and are subject to signal loss in areas with strong susceptibility gradients such as the tissue air surface close to the sinuses. The use of an ultra-short $T_E$ as in the current implementation will ameliorate this effect. A frequency selective pulse is further increasing the sensitivity due to $B_0$ offset. Areas with large resonance offset may lead to inaccurate $C_{B1}$ estimation due to partial excitation of PCr or signal bleeding from other metabolites. Future work could target this issue by incorporating $B_0$ maps to correct the $C_{B1}$ estimations.

The proposed method shares a similar limitation as other methods (DAM[32], dual-TR[35], AFI[15,26]), being sensitive to $T_1$ values. In this study, a uniform $T_1$ value was used based on values reported in the literature. Prior knowledge of tissue-specific $T_1$ values could enhance $C_{B1}$ mapping accuracy; however, such information is often unavailable. We therefore investigated the impact of $T_1$ on $C_{B1}$ accuracy using simulations and phantom results. In a rather rare and worst case ($C_{B1}$ = 0.1), the bias of $C_{B1}$ estimation was below 15% for a ± 25% difference in $T_1$. In a more likely scenario, phantom and simulation studies indicated that the $C_{B1}$ estimation bias was below 10% in a ± 25% range of $T_1$ ($C_{B1} = 0.81$). When $C_{B1}$ increases, estimation bias was less sensitive to $T_1$, e.g., below 10% in a ± 50% range of $T_1$ for $C_{B1} = 1.15$. This limitation can be further alleviated by increasing the $C_{B1}$ (i.e., using a higher transmit voltage), as the bias introduced by $T_1$ is much lower for higher $C_{B1}$ values (>1.25) as shown in Figure 4a. With the 8 ms Gaussian pulse used in this implementation, in vivo SAR limitations were far from being reached (2-3%), which allows the use of a higher transmit voltage. In studies with strong variation of $T_1$ (> 50%), a $T_1$ insensitive method such as BS-based methods[29,30,33] might be more favourable. The current fast readout sequence could be further adapted by including a Femi pulse for this purpose.

fDAM operates within SAR and peripheral nerve stimulation limits on a clinically approved UHF scanner. With the SAR limits far from being exceeded, the method could be considered even for nuclei with a lower gyromagnetic ratio. The strength of this method is the easy-to-implement look-up table approach. In combination with the fact that GRE sequences (or Spin Echo sequences) are widely available as standard sequences, fDAM is easy to use for other researchers or clinicians. To make the application more accessible, our sequence is available for Siemens XA60 and VB17A via c2p. The gradient readout can be easily changed using an external text file, and the sequence can be adapted to other body parts, nuclei, or readout trajectories. The reconstruction pipeline with the look-up table estimation is freely available on GitHub (https://github.com/MaSteWi/fDAM-B1-mapping.git). We expect that the proposed strategy along with recent advances in MR fingerprinting[12], novel RF coils[48] and denoising techniques[49,50], will contribute to significant reductions in total acquisition times making $^{31}$P MRSI more suitable for clinical settings.

## Conclusion

A fast short-$T_R$ double angle $B_1^+$ mapping method including a weighted stack of spiral trajectory and a frequency selective pulse was successfully implemented and demonstrated for $^{31}$P $B_1^+$ mapping in the human calf muscle and the brain. The novel method allows single subject $^{31}$P $C_{B1}$ mapping in both human calf muscle and brain in 10 min, with a spatial resolution of $14 \times 14 \times 20$ mm$^3$ and $7 \times 7 \times 20$ mm$^3$ respectively, showing promise for future applications in rapid X-nuclei imaging. This method can be applied to other fast imaging techniques available with the proposed look-up table approach.

**Author contribution**

MW: conceptualization, methodology, sequence development, experiments, data processing, writing original draft and review. AK: writing original draft and review. SB: experiments, data processing, writing original draft. YX and ZH: sequence development support and review of draft. YJ: spiral readout support and review of draft. DW: hardware support and review of draft. KP: support phantom creation and review of draft.  LX: conceptualization, methodology, writing original draft and review, supervision, resources, data curation, funding acquisition.

**Acknowledgment**

We acknowledge access to the facilities and expertise of the CIBM Center for Biomedical Imaging, a Swiss research center of excellence founded and supported by Lausanne University Hospital (CHUV), University of Lausanne (UNIL), Ecole Polytechnique Federale de Lausanne (EPFL), University of Geneva (UNIGE), and Geneva University Hospitals (HUG). Open access funding provided by EPFL.

**Funding information**

This work was supported by the Swiss National Science Foundation (grant no. 320030_189064 and 213769). Yun Jiang was partly supported by the NIH/NCI grant R37CA263583 and by Siemens Healthineers.

**Conflict of interest statement**

Yun Jiang receives research grant support from Siemens Healthineers.

**Figure Captions**

**Figure 1:** (a) The function between the actual FA and the estimated FA by cDAM is plotted for different $\sigma$ values. When using a small $\sigma$, cDAM develops an increasing bias in the FA estimation. The signal ratio r can be used to generate look up tables connecting it to a specific FA. In (b) look up tables for different $\sigma$ values are shown for fDAM. Values of the objective function $O$ over a range of $\alpha_1$ and $\sigma$ is shown in (c). The optimal $\alpha_1$ for a given $\sigma$ and vice-versa is indicated by the grey line. At the maximum value of the optimisation function, indicated by the white square, the optimal parameters $\alpha_{1,opt}$ and $\sigma_{opt}$ are found.

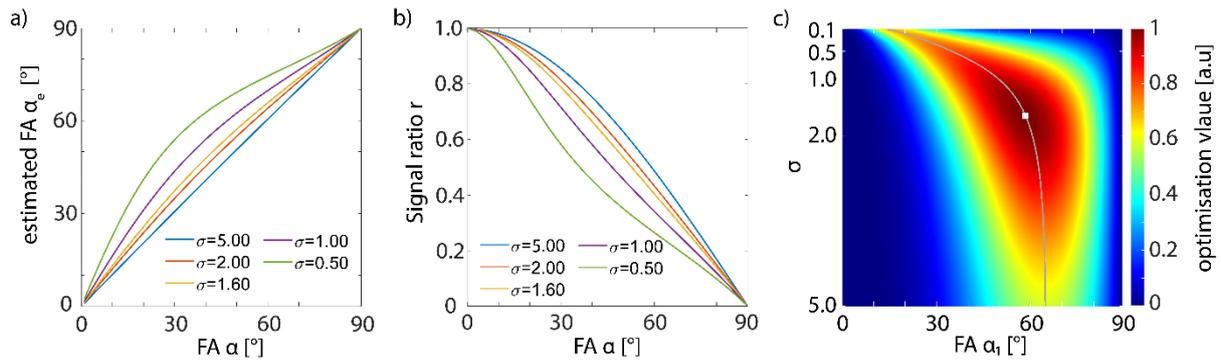

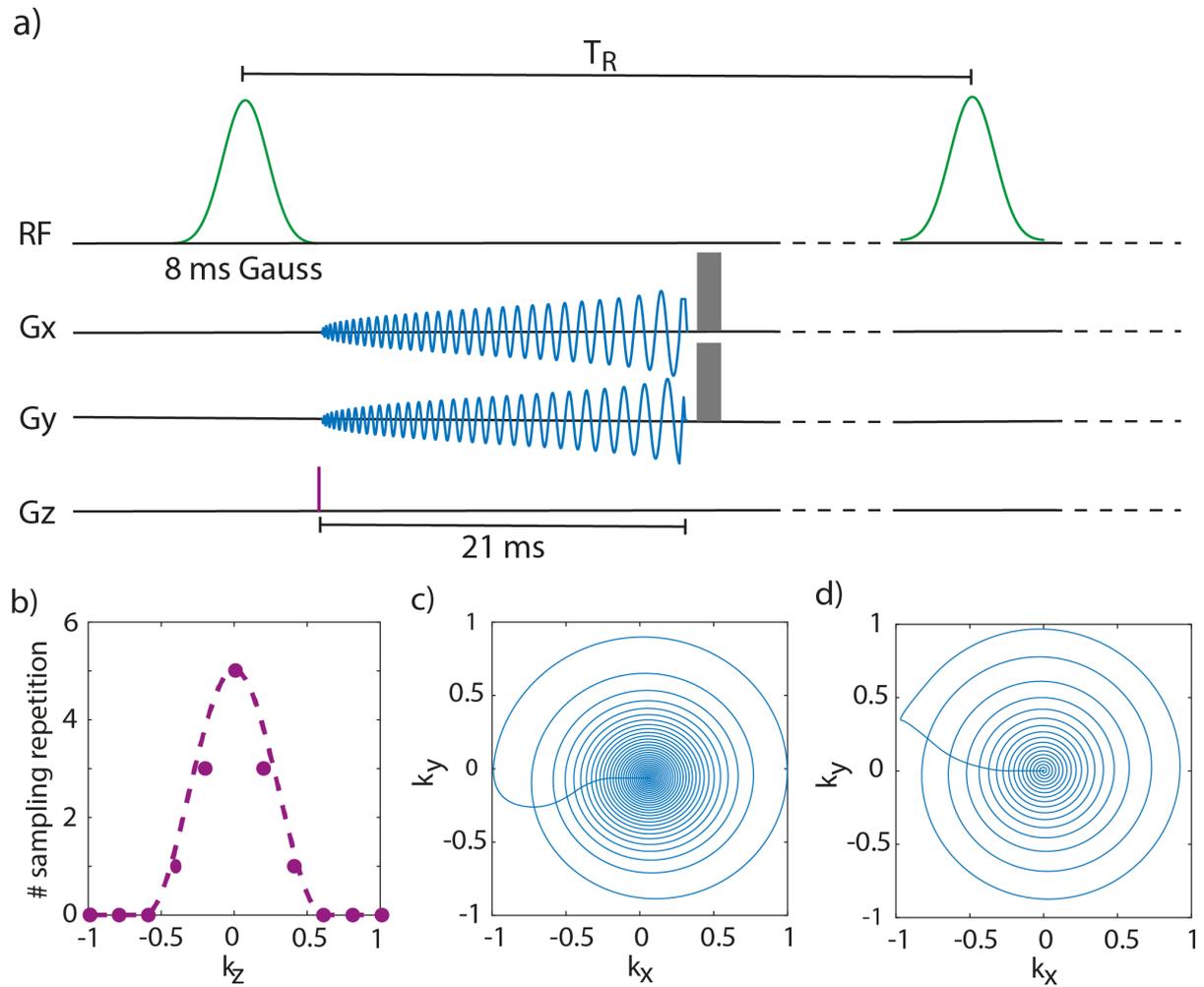

**Figure 2:** (a) Schematics of the $^{31}$P spiral GRE sequence. (b) The hamming weighted acquisition results with different numbers of averages per $k_z$ plain, (c) the spiral readout trajectory for the human calf muscle (VB17A: 21.08 ms, 10 µs sampling rate) and (d) brain (XA60: 18.72 ms, 10 µs sampling rate) experiment.

**Figure 3:** Comparison of $C_{B1}$ maps acquired by cDAM and fDAM in phantom experiments. (a) $C_{B1}$ maps of the centre slice from cDAM and fDAM of different $\sigma$ and (b) the difference maps with cDAM as a reference. (c) Correlation plot and (d) BA plot to compare $C_{B1}$ maps of fDAM with $\sigma_{opt}$=1.6 and cDAM over voxels of all slices. In (c) and (d), the colour bar shows the SNR of each voxel.

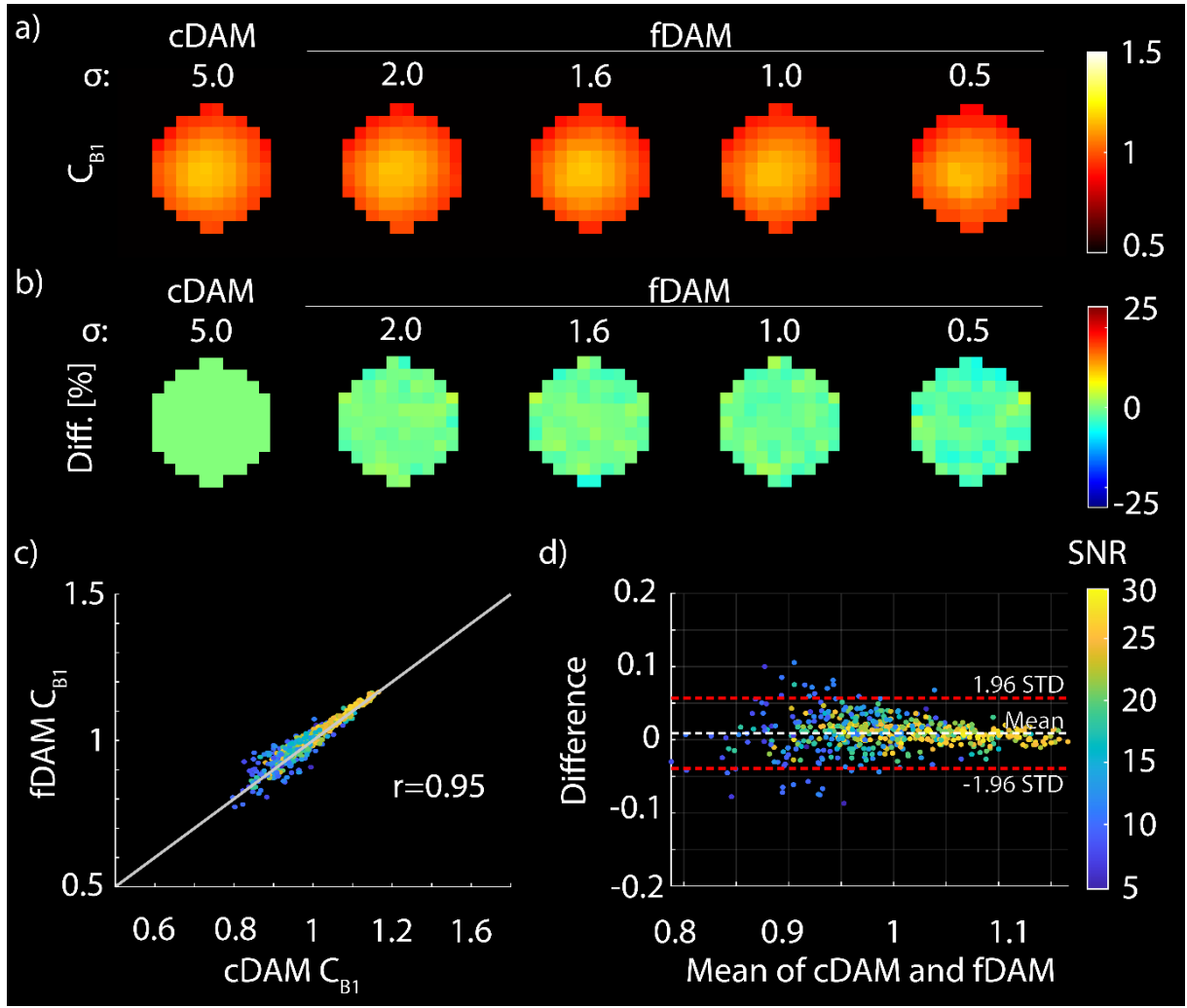

**Figure 4:** Evaluation of the introduced bias in $C_{B1}$ estimation due to $T_1$ deviation using the optimised $\alpha_{1,opt}$ and $\sigma_{opt}$. (a) The simulation results and (b) the phantom results for two different transmit voltages (175V and 250V) and the estimation error [%] when using deviated $T_1$ values (from -50% to 50%) in the look-up table. (c) Comparison of the phantom results for a voxel in the centre marked in orange (175V: $C_{B1}$=0.81) and blue (250V: $C_{B1}$ = 1.15) with the corresponding simulation results.

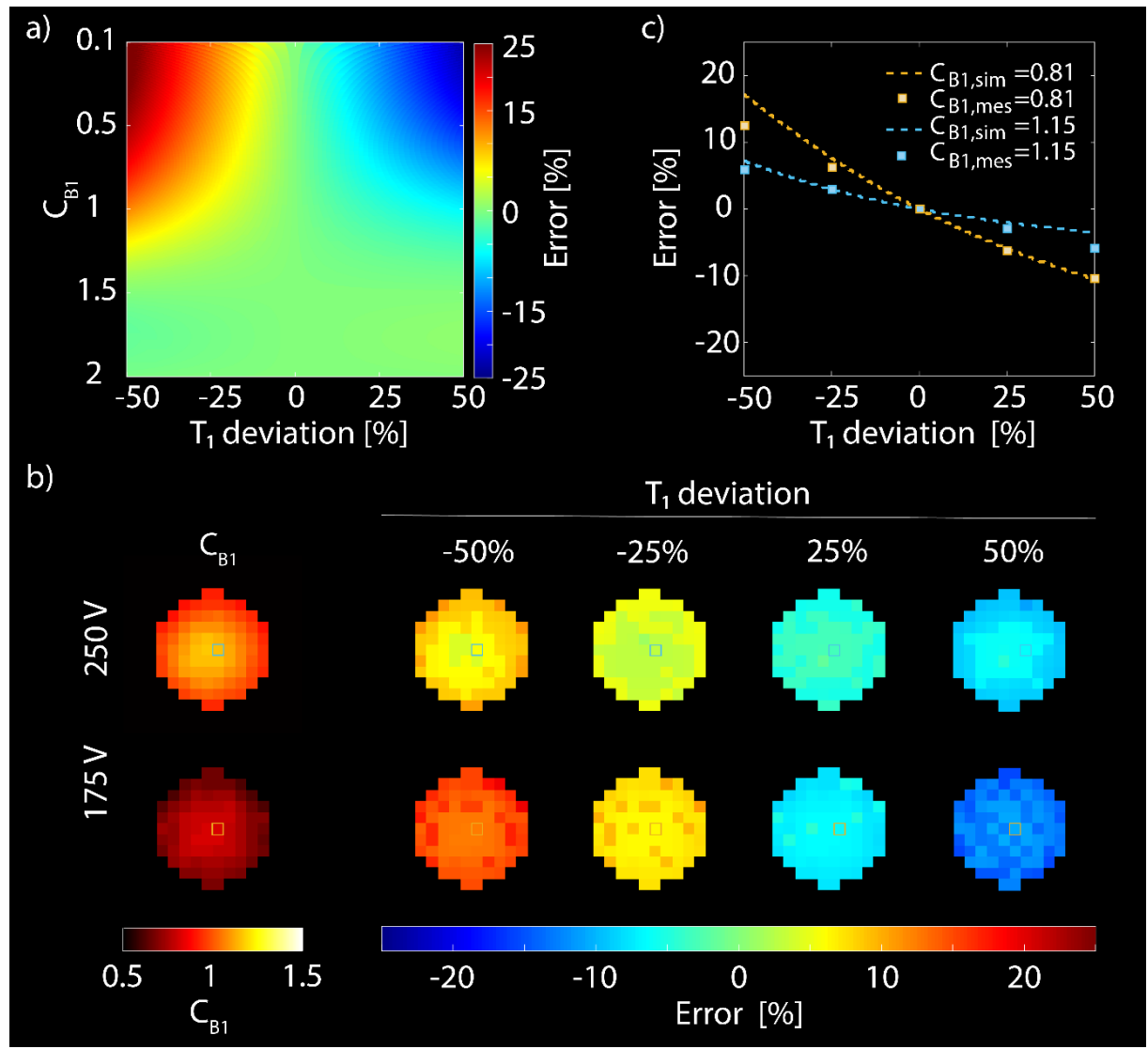

**Figure 5:** In vivo calf muscle images of participant 1 (a) and 2 (b), including the $^{1}$H GRE images, the $^{31}$P GRE1 images (for $\alpha_1$), the $^{31}$P GRE2 images (for $\alpha_2$), the $C_{B1}$ maps of cDAM and fDAM, as well as the difference maps (fDAM-cDAM)100/fDAM. Note that for the $C_{B1}$ and difference maps only voxels with SNR>3 are displayed.

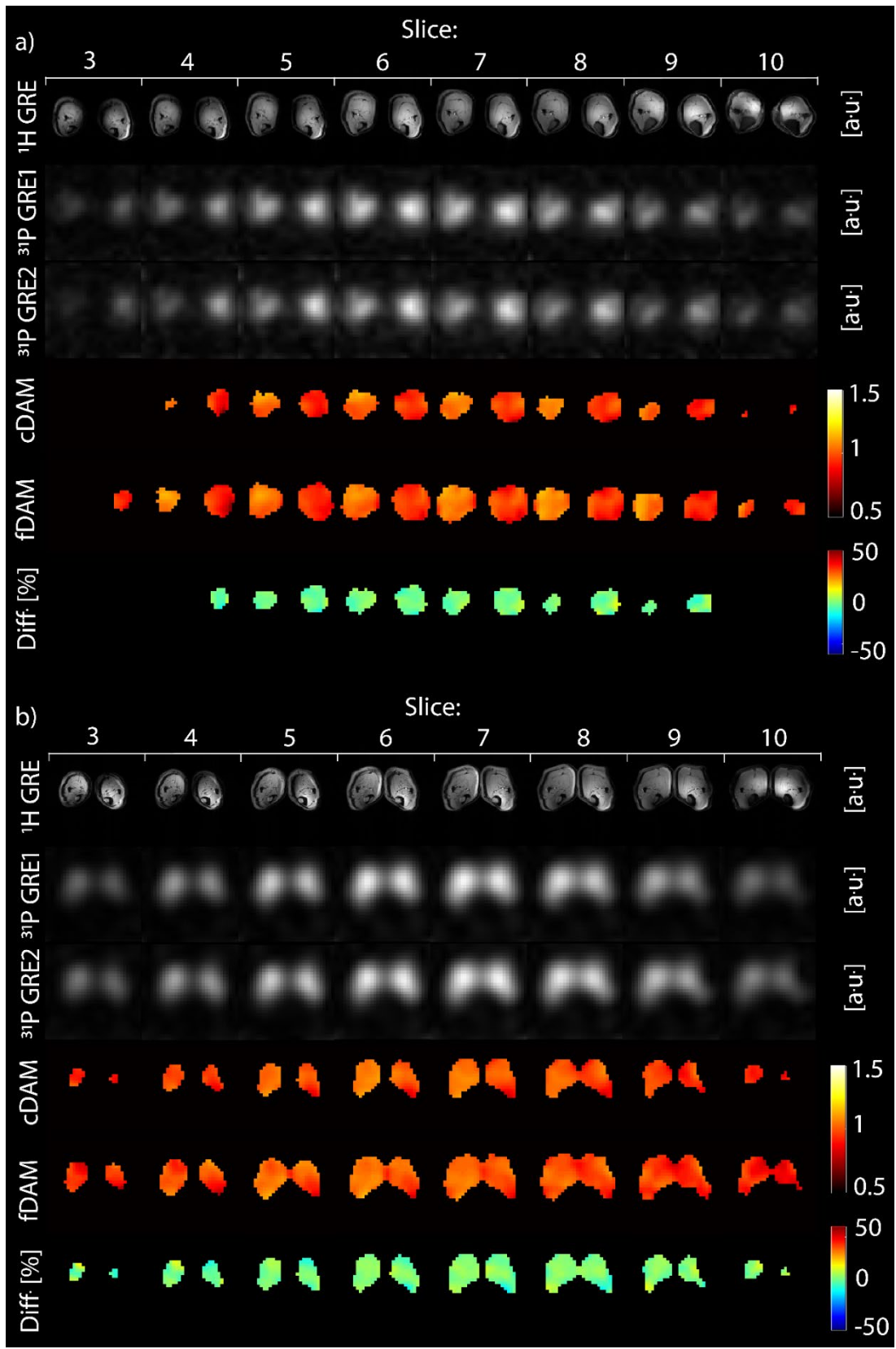

**Figure 6:** Comparison of $C_{B1}$ maps acquired by fDAM ($n_{\text{rep}} = 6$; 16 min) with those measured with fewer repetitions ($n_{\text{rep}} = 4$; 11 min and $n_{\text{rep}} = 4$; 6min) and their respective difference maps (participant 1 (a) and 2 (b)). The 3rd and 5th rows show the percentage difference maps of (Rep. 4 - Rep. 6)100/ Rep. 6 and (Rep. 2 - Rep. 6)100/ Rep. 6, respectively.

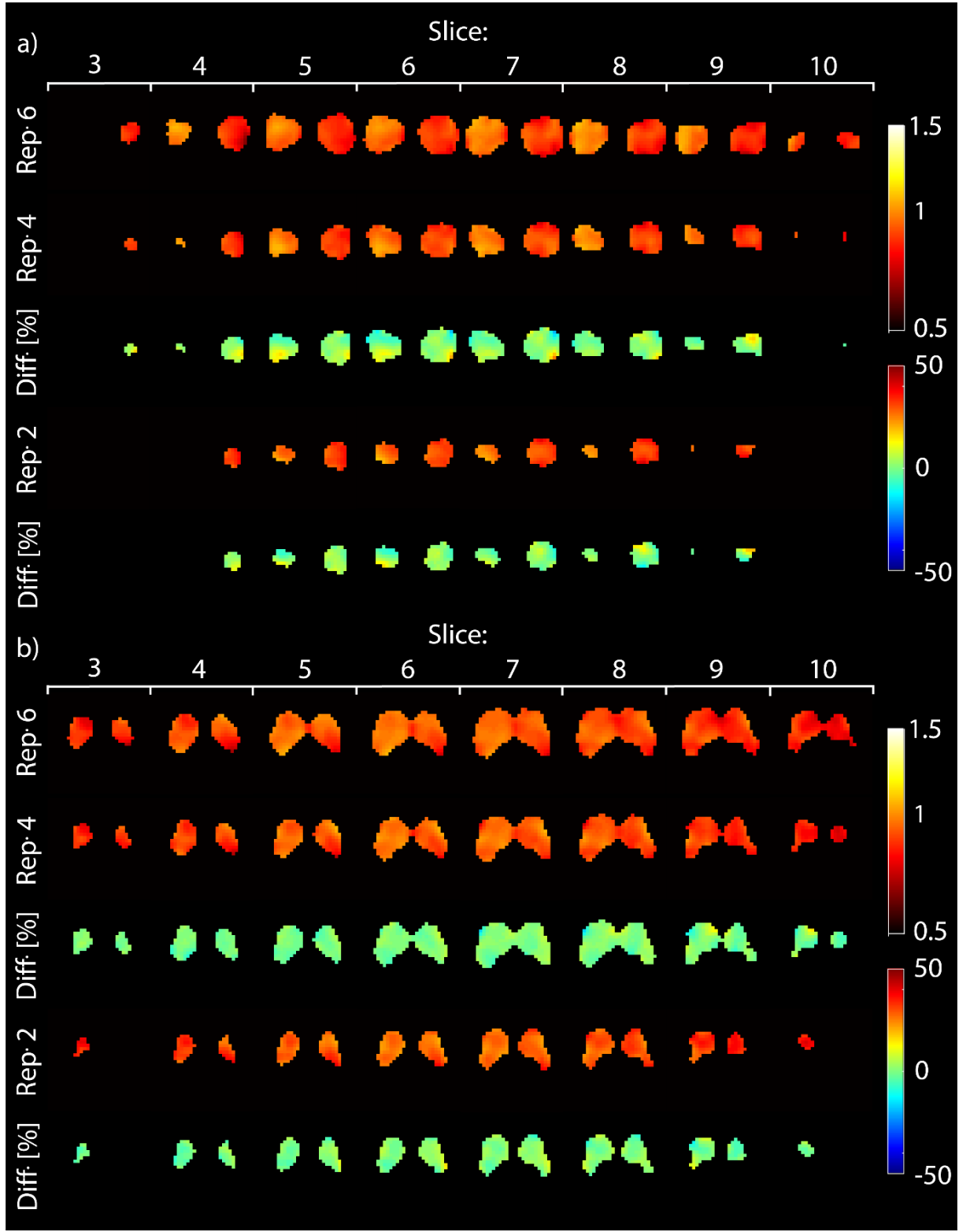

**Figure 7:** Comparison of $C_{B1}$ maps acquired by fDAM ($n_{\text{rep}} = 6$; 14:30 min) with those measured with less repetitions ($n_{\text{rep}} = 4$; 10:15 min and $n_{\text{rep}} = 2$; 5:30 min) and their respective difference maps. The reference $^1$H GRE images, the $^{31}$P GRE1 and GRE2 images with $n_{\text{rep}} = 6$ were shown on the top. The 6$^{\text{th}}$ row shows the percentage difference maps of (fDAM-cDAM)100/fDAM and the 8$^{\text{th}}$ and 10$^{\text{th}}$ rows show the percentage difference maps of (Rep. 4-Rep. 6)100/Rep. 6 and (Rep. 2-Rep. 6)100/Rep. 6, respectively.

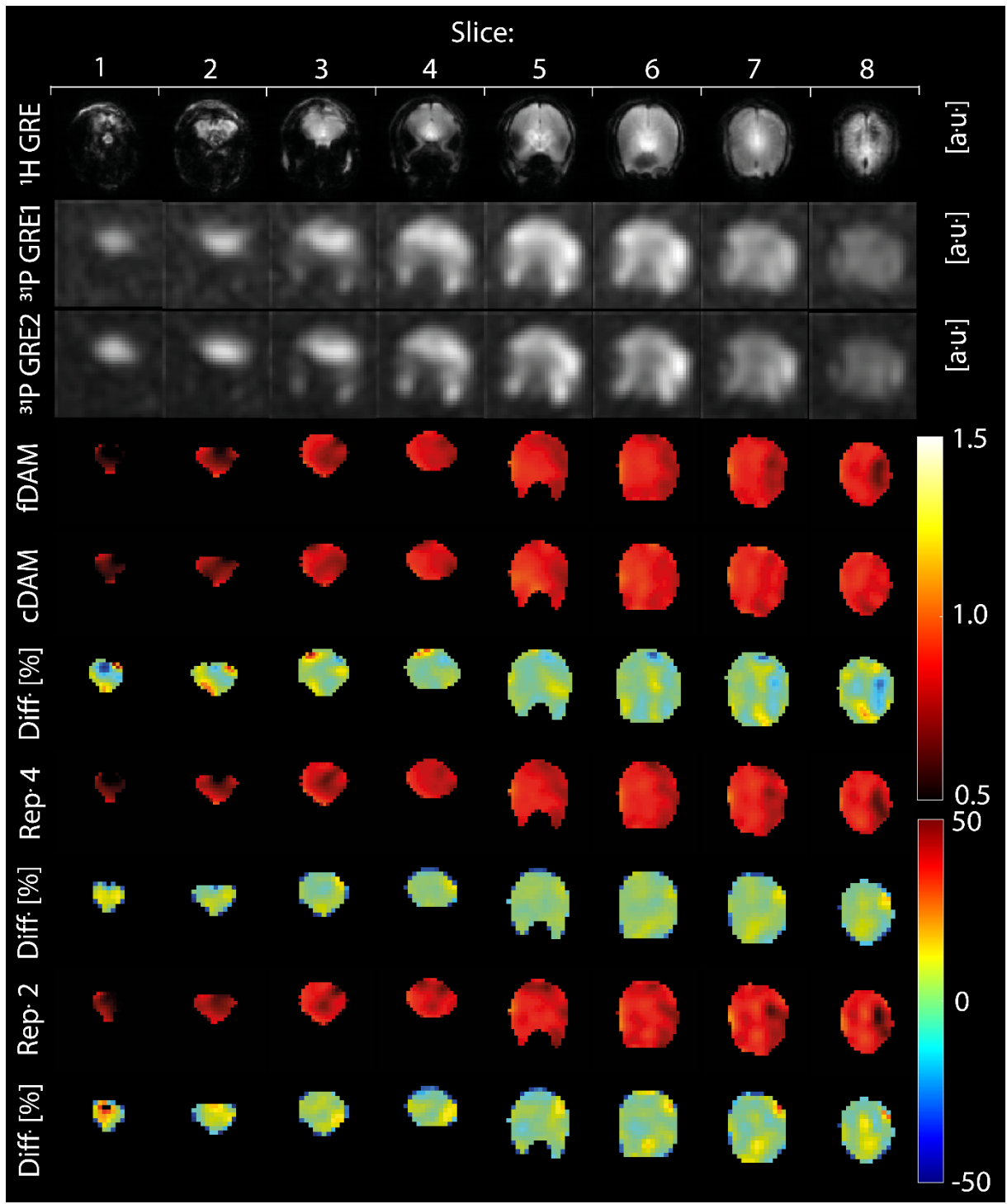

**References**

1. Huang Z, Gambarota G, Xiao Y, Wenz D, Xin L. Apparent diffusion coefficients of 31P metabolites in the human calf muscle at 7 T. *Magn Reson Mater Phys Biol Med*. 2023;36(2):309-315. doi:10.1007/s10334-023-01065-3

2. Xin L, Ipek Ö, Beaumont M, et al. Nutritional Ketosis Increases NAD+/NADH Ratio in Healthy Human Brain: An in Vivo Study by 31P-MRS. *Front Nutr*. 2018;5. Accessed July 1, 2022. https://www.frontiersin.org/article/10.3389/fnut.2018.00062

3. Klomp DWJ, van de Bank BL, Raaijmakers A, et al. 31P MRSI and 1H MRS at 7 T: initial results in human breast cancer. *NMR Biomed*. 2011;24(10):1337-1342. doi:10.1002/nbm.1696

4. Skupienski R, Steullet P, Do KQ, Xin L. Developmental changes in cerebral NAD and neuroenergetics of an antioxidant compromised mouse model of schizophrenia. *Transl Psychiatry*. 2023;13(1):1-9. doi:10.1038/s41398-023-02568-2

5. Forlenza OV, Wacker P, Nunes PV, et al. Reduced phospholipid breakdown in Alzheimer's brains: a 31P spectroscopy study. *Psychopharmacology (Berl)*. 2005;180(2):359-365. doi:10.1007/s00213-005-2168-8

6. Du F, Yuksel C, Chouinard VA, et al. Abnormalities in High-Energy Phosphate Metabolism in First-Episode Bipolar Disorder Measured Using 31P-Magnetic Resonance Spectroscopy. *Biol Psychiatry*. 2018;84(11):797-802. doi:10.1016/j.biopsych.2017.03.025

7. Du F, Cooper AJ, Thida T, et al. In Vivo Evidence for Cerebral Bioenergetic Abnormalities in Schizophrenia Measured Using 31P Magnetization Transfer Spectroscopy. *JAMA Psychiatry*. 2014;71(1):19-27. doi:10.1001/jamapsychiatry.2013.2287

8. Yuksel C, Chen X, Chouinard VA, et al. Abnormal Brain Bioenergetics in First-Episode Psychosis. *Schizophr Bull Open*. 2021;2(1):sgaa073. doi:10.1093/schizbullopen/sgaa073

9. Lei H, Zhu XH, Zhang XL, Ugurbil K, Chen W. In vivo 31P magnetic resonance spectroscopy of human brain at 7 T: An initial experience. *Magn Reson Med*. 2003;49(2):199-205. doi:10.1002/mrm.10379

10. Bogner W, Chmelik M, Schmid A i., Moser E, Trattnig S, Gruber S. Assessment of 31P relaxation times in the human calf muscle: A comparison between 3 T and 7 T in vivo. *Magn Reson Med*. 2009;62(3):574-582. doi:10.1002/mrm.22057

11. Valkovič L, Chmelík M, Just Kukurova I, et al. Time-resolved phosphorous magnetization transfer of the human calf muscle at 3T and 7T: A feasibility study. *Eur J Radiol*. 2013;82(5):745-751. doi:10.1016/j.ejrad.2011.09.024

12. Widmaier M, Lim SI, Wenz D, Xin L. Fast in vivo assay of creatine kinase activity in the human brain by 31P magnetic resonance fingerprinting. *NMR Biomed*. 2023;36(11):e4998. doi:10.1002/nbm.4998

13. Yang QX, Wang J, Zhang X, et al. Analysis of wave behavior in lossy dielectric samples at high field. *Magn Reson Med*. 2002;47(5):982-989. doi:10.1002/mrm.10137

14. Tofts PS. Standing Waves in Uniform Water Phantoms. *J Magn Reson B*. 1994;104(2):143-147. doi:10.1006/jmrb.1994.1067

15. Parasoglou P, Xia D, Chang G, Regatte RR. 3D-mapping of phosphocreatine concentration in the human calf muscle at 7 T: Comparison to 3 T. *Magn Reson Med*. 2013;70(6):1619-1625. doi:10.1002/mrm.24616

16. Kemp GJ, Meyerspeer M, Moser E. Absolute quantification of phosphorus metabolite concentrations in human muscle in vivo by 31P MRS: a quantitative review. *NMR Biomed*. 2007;20(6):555-565. doi:10.1002/nbm.1192

17. Buchli R, Boesiger P. Comparison of methods for the determination of absolute metabolite concentrations in human muscles by 31P MRS. *Magn Reson Med*. 1993;30(5):552-558. doi:10.1002/mrm.1910300505

18. Buchli R, Martin E, Boesiger P. Comparison of calibration strategies for the in vivo determination of absolute metabolite concentrations in the human brain by 31P MRS. *NMR Biomed*. 1994;7(5):225-230. doi:10.1002/nbm.1940070505

19. Dunn JF, Kemp GJ, Radda GK. Depth selective quantification of phosphorus metabolites in human calf muscle. *NMR Biomed*. 1992;5(3):154-160. doi:10.1002/nbm.1940050309

20. Doyle VL, Payne GS, Collins DJ, Verrill MW, Leach MO. Quantification of phosphorus metabolites in human calf muscle and soft-tissue tumours from localized MR spectra acquired using surface coils. *Phys Med Biol*. 1997;42(4):691. doi:10.1088/0031-9155/42/4/006

21. Bottomley PA, Atalar E, Weiss RG. Human cardiac high-energy phosphate metabolite concentrations by 1D-resolved NMR spectroscopy. *Magn Reson Med*. 1996;35(5):664-670. doi:10.1002/mrm.1910350507

22. Insko EK, Bolinger L. Mapping of the Radiofrequency Field. *J Magn Reson A*. 1993;103(1):82-85. doi:10.1006/jmra.1993.1133

23. Pohmann R, Scheffler K. A theoretical and experimental comparison of different techniques for B1 mapping at very high fields. *NMR Biomed*. 2013;26(3):265-275. doi:10.1002/nbm.2844

24. Eggenschwiler F, Kober T, Magill AW, Gruetter R, Marques JP. SA2RAGE: A new sequence for fast B1+-mapping. *Magn Reson Med*. 2012;67(6):1609-1619. doi:10.1002/mrm.23145

25. Bouhrara M, Spencer RG. Steady State Double Angle Method for Rapid B1 Mapping. *Magn Reson Med*. 2019;82(1):189-201. doi:10.1002/mrm.27708

26. Yarnykh VL. Actual flip-angle imaging in the pulsed steady state: A method for rapid three-dimensional mapping of the transmitted radiofrequency field. *Magn Reson Med*. 2007;57(1):192-200. doi:10.1002/mrm.21120

27. Morrell GR. A phase-sensitive method of flip angle mapping. *Magn Reson Med*. 2008;60(4):889-894. doi:10.1002/mrm.21729

28. Jy P. B_1 mapping using phase information created by frequency-modulated pulses. *Proc 16th Annu Meet ISMRM Tor Can 2008*. 2008;361. Accessed November 14, 2023. https://cir.nii.ac.jp/crid/1574231875896909056

29. Bloch F, Siegert A. Magnetic Resonance for Nonrotating Fields. *Phys Rev*. 1940;57(6):522-527. doi:10.1103/PhysRev.57.522

30. Sacolick LI, Wiesinger F, Hancu I, Vogel MW. B1 mapping by Bloch-Siegert shift. *Magn Reson Med*. 2010;63(5):1315-1322. doi:10.1002/mrm.22357

31. Khalighi MM, Rutt BK, Kerr AB. Adiabatic RF pulse design for Bloch-Siegert B mapping. *Magn Reson Med*. 2013;70(3):829-835. doi:10.1002/mrm.24507

32. Greenman RL, Rakow-Penner R. Evaluation of the RF field uniformity of a double-tuned 31P/1H birdcage RF coil for spin-echo MRI/MRS of the diabetic foot. *J Magn Reson Imaging*. 2005;22(3):427-432. doi:10.1002/jmri.20372

33. Clarke WT, Robson MD, Rodgers CT. Bloch-Siegert -mapping for human cardiac 31P-MRS at 7 Tesla. *Magn Reson Med*. 2016;76(4):1047-1058. doi:10.1002/mrm.26005

34. Ishimori Y, Shimanuki T, Kobayashi T, Monma M. Fast B1 Mapping Based on Double-Angle Method with T1 Correction Using Standard Pulse Sequence. *J Med Phys*. 2022;47(1):93-98. doi:10.4103/jmp.jmp_78_21

35. Chmelík M, Považan M, Jírů F, et al. Flip-angle mapping of 31P coils by steady-state MR spectroscopic imaging. *J Magn Reson Imaging*. 2014;40(2):391-397. doi:10.1002/jmri.24401

36. Lee JH, Hargreaves BA, Hu BS, Nishimura DG. Fast 3D imaging using variable-density spiral trajectories with applications to limb perfusion. *Magn Reson Med*. 2003;50(6):1276-1285. doi:10.1002/mrm.10644

37. Rasche V, Proksa R, Sinkus R, Bornert P, Eggers H. Resampling of data between arbitrary grids using convolution interpolation. *IEEE Trans Med Imaging*. 1999;18(5):385-392. doi:10.1109/42.774166

38. Aurenhammer F. Voronoi diagrams—a survey of a fundamental geometric data structure. *ACM Comput Surv*. 1991;23(3):345-405. doi:10.1145/116873.116880

39. Meng Sang Ong. Arbitrary Square Bounded Voronoi Diagram. MATLAB Central File Exchange. Published January 19, 2024. Accessed January 19, 2024. https://www.mathworks.com/matlabcentral/fileexchange/30353-arbitrary-square-bounded-voronoi-diagram

40. Jackson JI, Meyer CH, Nishimura DG, Macovski A. Selection of a convolution function for Fourier inversion using gridding (computerised tomography application). *IEEE Trans Med Imaging*. 1991;10(3):473-478. doi:10.1109/42.97598

41. Beatty PJ, Nishimura DG, Pauly JM. Rapid gridding reconstruction with a minimal oversampling ratio. *IEEE Trans Med Imaging*. 2005;24(6):799-808. doi:10.1109/TMI.2005.848376

42. Walsh DO, Gmitro AF, Marcellin MW. Adaptive reconstruction of phased array MR imagery. *Magn Reson Med*. 2000;43(5):682-690. doi:10.1002/(sici)1522-2594(200005)43:5<682::aid-mrm10>3.0.co;2-g

43. Determination of Signal-to-Noise Ratio (SNR) in Diagnostic Magnetic Resonance Imaging. NEMA. Published April 30, 2015. Accessed January 4, 2024. https://www.nema.org/Standards/view/Determination-of-Signal-to-Noise-Ratio-in-Diagnostic-Magnetic-Resonance-Imaging

44. Ren J, Sherry AD, Malloy CR. 31P-MRS of healthy human brain: ATP synthesis, metabolite concentrations, pH, and T1 relaxation times. *NMR Biomed*. 2015;28(11):1455-1462. doi:10.1002/nbm.3384

45. Tomanek B, Volotovskyy V, Gruwel MLH, McKenzie E, King SB. Double-frequency birdcage volume coils for 4.7T and 7T. *Concepts Magn Reson Part B Magn Reson Eng*. 2005;26B(1):16-22. doi:10.1002/cmr.b.20038

46. Valkovič L, Dragonu I, Almujayyaz S, et al. Using a whole-body 31P birdcage transmit coil and 16-element receive array for human cardiac metabolic imaging at 7T. *PLoS ONE*. 2017;12(10):e0187153. doi:10.1371/journal.pone.0187153

47. Wang J, Mao W, Qiu M, Smith MB, Constable RT. Factors influencing flip angle mapping in MRI: RF pulse shape, slice-select gradients, off-resonance excitation, and B0 inhomogeneities. *Magn Reson Med*. 2006;56(2):463-468. doi:10.1002/mrm.20947

48. D. Wenz, T. Dardano, M. Widmaier, S. Lim, Z. Huang, L. Xin. Degenerate birdcage coil mixed with bent dipole antennas to enhance central SNR in phosphorus MRI/MRS of human brain at 7T. In: *Proc. Intl. Soc. Mag. Reson. Med. 31 (2023)*. ; 2023.

49. Lustig M, Donoho DL, Santos JM, Pauly JM. Compressed Sensing MRI. *IEEE Signal Process Mag*. 2008;25(2):72-82. doi:10.1109/MSP.2007.914728

50. Mosso J, Simicic D, Şimşek K, Kreis R, Cudalbu C, Jelescu IO. MP-PCA denoising for diffusion MRS data: promises and pitfalls. *NeuroImage*. 2022;263:119634. doi:10.1016/j.neuroimage.2022.119634

# Supplementary Files

## Figure S1:

Monte Carlo simulations were conducted for a fDAM setup assuming $T_1$=3.4 s, $T_R$=5.4 s and $α_1$=59°. 61 GRE signal pairs were simulated (Eq. 2) and ranging for a $C_{B1}$ value from 0.7 to 1.3. Each signal pair was manifolded by 1000 and random Gaussian noise was added before $C_{B1}$ estimation. The process was repeated for 20 signal-to-noise ratios (SNRs) ranging from 1 to 20. The standard deviation and mean error over all $C_{B1}$ values and noise manifolds were displayed. The mean error deviates above 10% for SNR<3. Therefore, SNR>3 was chosen as a threshold.

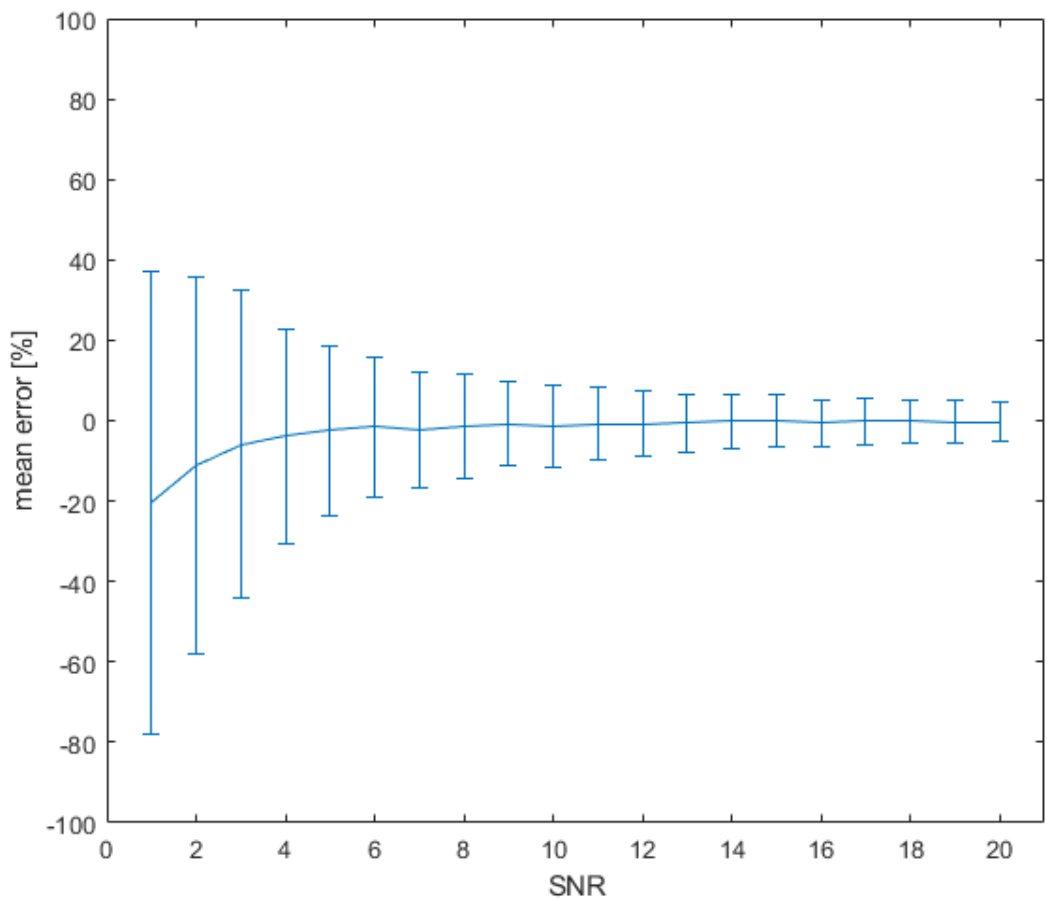

**Figure S2:**

$B_1^+$-histograms in [µT] for all 3 experiments: (a) Participant 1 calf muscle, (b) Participant 2 calf muscle, (c) Participant 3 brain.

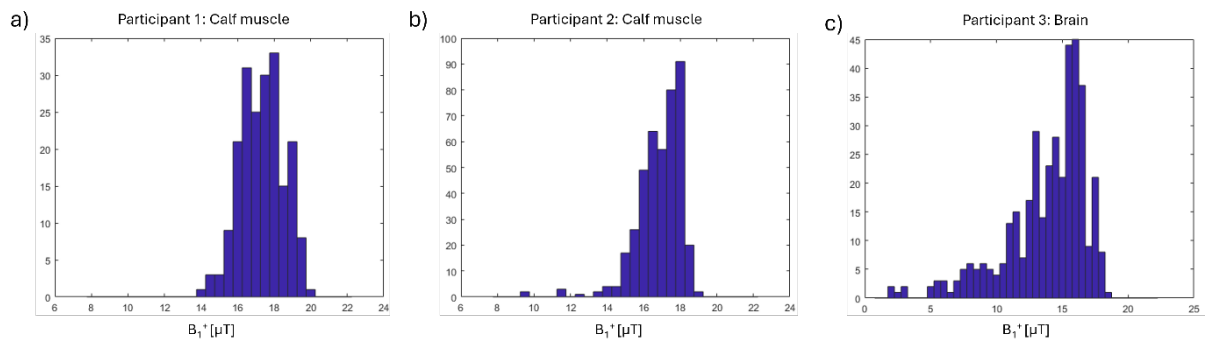